# Transition Metal Dimers and Physical Limits on Magnetic Anisotropy

Tor O. Strandberg[1], Carlo. M. Canali[1] & Allan H. MacDonald[2]

[1]*Division of Physics, Department of Natural Sciences, Kalmar University, S31912 Kalmar, Sweden.*

[2]*Department of Physics, University of Texas at Austin, Austin TX 78712, USA.*

**Recent advances in nanoscience have raised interest in the minimum bit size required for classical information storage, *i.e.* for bistability with suppressed quantum tunnelling and energy barriers that exceed ambient temperatures. In the case of magnetic information storage much attention has centred on molecular magnets[1] with bits consisting of ~ 100 atoms, magnetic uniaxial anisotropy energy barriers ~ 50 K, and very slow relaxation at low temperatures. In this article we draw attention to the remarkable magnetic properties of some transition metal dimers which have energy barriers approaching ~ 500 K with only two atoms. The spin dynamics of these ultra small nanomagnets is strongly affected by a Berry phase which arises from quasi-degeneracies at the electronic Highest Occupied Molecular Orbital (HOMO) energy. In a giant spin-approximation, this Berry phase makes the effective reversal barrier thicker.**

Interest[2] in magnetic nanoparticles, which typically contain tens of thousands of magnetic atoms, has been spurred both by the crucial role that they play in advanced magnetic information storage devices, and by the light that investigating magnetism at the nanoscale sheds on the fundamental interactions responsible for the magnetic state. As the frontier advances, interest is shifting to still smaller size scales. Dimers represent the small size end point in the transition metal cluster crossover from nanoparticle[2] to molecular magnetic properties. The magnetic moments per atom in clusters of the 3d transition metal elements Fe, Co, and Ni are typically enhanced[3] compared to those of the corresponding bulk metals, while 4d elements like Rh and Pd, which are not magnetic in bulk, can display a magnetic moment in small clusters. Cluster research has progressed rapidly in the last decade, partially because of experimental advances[4,5] in synthesis and characterization. The key property of any magnetic nanoparticle or cluster is its magnetic anisotropy energy, which is normally defined in classical terms as a dependence of energy on magnetization direction ($\hat{n}$). When $\hat{n}$ can be treated as a classical variable, it will be trapped along an easy magnetization direction whenever the thermal energy $k_B T$ is much smaller than the anisotropy energy. Since the exchange interaction responsible for magnetism is rotationally invariant, the anisotropy energy comes from spin-orbit interactions.

In discussing how the magnetic anisotropy of a transition metal behaves at the nanoscale and below, it is useful to consider two extreme limits. In individual atoms, the magnetic state is determined by Hund's rules, giving rise to large spin and orbital moments. There is no anisotropy energy in a single atom because it is invariant under simultaneous spin and orbital rotation, even when spin-orbit coupling is included. In the other extreme - bulk systems - the spin moment per atom is substantially reduced from its atomic value, to zero in paramagnetic metals. When spin-orbit interactions are included $\hat{n}$ is coupled to the atomic arrangement. The variation of magnetic properties between bulk and atomic limits is presently the subject of intense study, both experimentally[6] and theoretically[7].

As we explain below, transition metal dimers occupy a very special place in this landscape because they are rotationally invariant around one axis only - the molecular axis. It follows that magnetic anisotropy can appear already at first order when spin-orbit interactions are treated perturbatively, and can therefore be anomalously strong. We first discuss in a qualitative way the delicate balances that determine which dimers have this giant magnetic anisotropy, using a simple bonding orbital language description of *ab initio* Spin Density Functional Theory (SDFT) calculations performed using VASP[8], and then turn our attention to the size of the anisotropy energy and the way in which it is quantized.

Transition metals have occupied s ($m_\ell$=0) and d ($m_\ell$=0, ±1, ±2) valence orbitals. In a dimer, an orbital centred on one atom which has azimuthal angular momentum $m_\ell$ along the molecular axis hybridizes with orbitals on the other atom that have the same $m_\ell$, leading to bonding and antibonding combinations. In a DFT description, the exchange potential lowers the energies of majority spin orbitals compared to those of minority spin orbitals. (The DFT Kohn-Sham orbital energies of $Co_2$ and $Ni_2$ with and without spin-orbit coupling are illustrated in Fig. 1a and b, labelled using standard notation[9].) For the late transition elements on which we focus, all 12 majority spin orbitals, except the s-like $m_\ell$=0 antibonding orbital, are normally occupied (Ru is one exception). This leaves 5 minority spin orbitals ($S$=3) in Fe, 7 ($S$=2) in the Co column, and 9 ($S$=1) in the Ni column. The energetic ordering of the d-like minority orbitals usually follows the pattern σ, π, δ, δ$^*$, π$^*$, σ, where σ refers to $m_\ell$=0 orbitals, π to degenerate $m_\ell$=±1 orbitals, and δ to degenerate $m_\ell$=±2 orbitals, and the superscript $^*$ indicates the antibonding combination. (This rule for ordering energy levels does not hold as well for the majority spin because of stronger orbital dependence in the exchange energy). The energy of the s-like σ orbital tends to decrease relative to the d-like orbitals, when moving up or to the left in the periodic table, whereas its σ$^*$ counterpart is always highest in energy.

Giant magnetic anisotropy in the dimers occurs when a singly occupied HOMO has a two-fold orbital degeneracy in the absence of spin-orbit coupling, implying a ground state with angular momentum along the z-axis equal to ±$m_\ell$. In $Co_2$ both d-like and s-like σ orbitals are occupied along with π and δ orbitals, leaving a doubly-degenerate, singly occupied $\delta^*$ HOMO. In the Ni dimer on the other hand, both $\delta^*$ minority orbitals are occupied and the $\sigma^*$ orbital exceptionally falls below the $\pi^*$, implying no giant anisotropy. In $Fe_2$ the $\delta$ and d-like $\sigma^*$-orbitals compete narrowly for HOMO status, complicating the anisotropy energy calculation leading ultimately to reduced anisotropy. The 4d elements Rh and Pd behave like their 3d counterparts, Co and Ni. In Ru, stronger bonding compared to Fe causes both s- and d-like $\sigma^*$ orbitals to be unoccupied, so that $S=2$ in the absence of spin-orbit coupling, with a doubly degenerate, but doubly occupied $\pi$ majority HOMO. Only $Co_2$ and $Rh_2$ satisfy the requirements for giant anisotropy.

Our SDFT calculation results, summarized in Table 1, are in good agreement with previous DFT and configuration-interaction calculations[10-13] and with available experimental[14,15] data. We caution the reader however, that the balance between orbital dependent bonding and exchange energies which determines the character of the ground state is extremely delicate when the level structure near the Fermi energy is dense, making reliable theoretical predictions difficult for some dimers. In $Ir_2$, for example, our SDFT calculations were unable to reliably resolve a close competition between a higher spin state at larger bond length and a lower spin state at a smaller bond length.

We now turn our attention to the anisotropy energies of $Co_2$ and $Rh_2$. To understand the large uniaxial anisotropies estimated by SDFT, illustrated in Fig. 1c and d, it is instructive to examine the influence of spin-orbit interaction on the individual molecular orbitals, illustrated in the right-hand panels of Fig. 1a and b. Up to double-counting corrections, the total anisotropy energy is simply the sum of the spin orientation dependent shifts in occupied orbital energies[16]. The first column of levels in the right-hand panels of Fig. 1a and b, illustrates level shifts when the total magnetic moment of the dimer is along a direction $x$ in the plane perpendicular to the dimer axis. The level shifts produced by spin-orbit interactions are small compared to the scale $\xi_d$ of the d-shell spin-orbit interaction ($H_{SO} = \xi_d \, \mathbf{s} \cdot \mathbf{L}$ with $\xi_d \sim 85$ meV in Co and $\sim 140$ meV in Rh) because the expectation value of $H_{SO}$ is zero for all orbitals. The unperturbed states are products of spins polarized along the $\pm x$ direction and orbital states for which the expectation value of $L_x$ is zero. In every case the shifts arise at higher order in perturbation theory and are at most $\sim \xi_d^2 / W_d \sim 0.1 \xi_d$, where $W_d$ is the typical d-orbital bonding energy. However, when the magnetic moment points along the dimer axis $z$, the first order spin-orbit interaction matrix elements are $\pm \xi_d \left\langle |L_z| \right\rangle / 2 = \pm m_\ell \, \xi_d / 2$. When a $m_\ell \neq 0$ HOMO level is singly occupied, only the lower energy member of the doublet is occupied, and the level makes a first order contribution to the anisotropy energy equal to $m_\ell \, \xi_d / 2$. For $Co_2$ and $Rh_2$ this first order anisotropy energy ($\xi_d$) is larger than the full anisotropy energy in Fig. 1c and d. Because of the symmetry of our problem, the full anisotropy energy is an even function of $\cos(\theta)$. When the anisotropy energy is evaluated at first order in perturbation theory, it is proportional to $|\cos(\theta)|$, i.e. it is a non-analytic function of $\cos(\theta)$. This non-analytic behaviour, which is caused by the level crossing occurring at $\theta = \pi/2$ only in systems without spin-orbit coupling, is an artefact of truncating at first-order in perturbation theory. The cusp at $\theta = \pi/2$ is always rounded out by higher-order terms and the anisotropy energy is in fact an *even* analytic function of $\cos(\theta)$. The degree of rounding is a complex issue of many-body physics, which when estimated by DFT calculations, is influenced by how the HOMO-LUMO occupancies are handled through the `so-called' smearing parameter. (See Supplementary Information). The full SDFT anisotropy energy is $\sim 30$ meV in $Co_2$ and $\sim 45$ meV in $Rh_2$. The anisotropy energy of a dimer with a non-degenerate or a doubly occupied and doubly degenerate HOMO - *e.g.* that of $Ni_2$ displayed in Fig. 1e - is typically one order of magnitude smaller. Both $Ni_2$ and $Pd_2$ have small easy-plane anisotropies.

Nanomagnets and molecular magnets are often described at low energies using a giant spin approximation in which the Hamiltonian has a single spin degree of freedom representing the collective magnetization of the system. Theories of vibration or hyperfine interaction relaxation between different magnetic states start[1] from a quantum Hamiltonian for the isolated giant spin, **J**. In the case of molecular magnets the appropriate Hamiltonian can be derived starting from a more microscopic model in which individual spin degrees of freedom are associated with particular magnetic ions. This approach does not work in metal clusters because the spins of itinerant electrons are tied to molecular orbitals and not to individual atoms. In addition, because of the delicate balance between bonding and exchange splittings discussed above, the combination of orbitals that contributes to the electronic state will vary with $\hat{n}$ once spin-orbit interactions are included. To overcome these obstacles, we derive approximate giant spin effective Hamiltonians for transition metal dimers by following the approach suggested in Ref. 17 in which the effective Hamiltonian $H(\mathbf{J})$ is obtained by integrating out electronic degrees of freedom which are presumed to be fast. (See Supplementary Information). The effective theory contains a remnant of the fast degrees of freedom in the form of a Berry phase term and associated Berry curvature, $C[\hat{n}]$, which describe the topologically non-trivial dependence of the many-body wavefunction on $\hat{n}$. The effective giant spin of this theory is a so-called Chern number $J$, a topological invariant equal to the average of $C[\hat{n}]$ over all directions $\hat{n}$, which can only take on values equal to multiples of half-integers. $J$ includes the effect of spin-orbit interactions in an exact manner. Note that in the usual derivations of effective spin Hamiltonians for magnetic clusters[20] and molecular magnets[21] the giant spin is normally chosen in an *ad hoc* fashion, for example by taking it as equal to the half-integer value nearest to the total magnetic moment or to the total spin in the absence of spin-orbit coupling. Then the Hamiltonian is simply obtained by quantizing directly the anisotropy energy function $E[\hat{n}]$ in the Hilbert space of dimension $2S+1$, without taking into account the effect of the Berry curvature. As we show below, the Berry phase can qualitatively change the functional form of the Hamiltonian.

Table 1 shows the Chern number $J$ of the Ground State (GS) calculated with spin-orbit coupling, compared to the total spin $S$ in the absence of spin-orbit coupling for several transition-metal dimers. The Chern number $J$ differs from $S$ only in systems with HOMO degeneracy. The difference between $J$ and $S$ can be understood in terms of the Berry phase associated with the avoided crossing of molecular orbitals with $L_z=\pm m$; since the $\mathbf{s}\cdot\mathbf{L}$ coupling appears at order $2m$ in perturbation theory it is proportional to $e^{2im\phi}$ where $\phi$ is the azimuthal spin-orientation. The results for Co and Rh can be understood intuitively as indicating that the spin and orbital moments add in the ground state, as expected when a minority spin orbital is at the Fermi level.

In Fig. 1c and d, we have plotted the anisotropy energy $E[\hat{n}]$ and the classical giant-spin Hamiltonian $\mathcal{H}[J,\hat{n}]$ for $Co_2$ and $Rh_2$ as a function of the angle $\theta$ between $\hat{n}$ and the dimer axis. The function $\mathcal{H}[J,\hat{n}]$ can be interpreted as an effective magnetic anisotropy energy for the giant spin, which includes the effect of the Berry curvature $C[\hat{n}]$. The maximum of $E[\hat{n}]$ in the equatorial plane of moment orientations ($\theta = \pi/2$) is replaced in $\mathcal{H}[J,\hat{n}]$ by a flat plateau, stretching over a large range of polar orientations. The net result is a thicker anisotropy energy barrier between the minima at the uniaxial directions $\theta = 0, \pi$. A steeper increase of the energy away from the poles implies that the oscillation frequencies of the collective magnetization around its equilibrium position are stiffer than the ones predicted by treatments that fail to include Berry phase corrections. This effect is partially analogous to the break-down of the Born-Oppenheimer approximation for the electron-phonon interaction in molecular systems. It has been recognized recently[22] that standard electronic structure calculations of polyatomic systems can yield faulty global minima configurations and vibrational frequencies, if Berry curvature effects associated with closely avoided level crossings are neglected. When $J = S$ the giant spin Hamiltonian is essentially indistinguishable from the original anisotropy function $E[\hat{n}]$, as shown in Fig. 1d . For these cases the formalism described in the present paper puts the commonly employed *ad hoc* recipes on a firmer footing.

The modified anisotropy energy landscape can have important consequences also for the large amplitude collective dynamics involved in thermal reversal processes, and for quantum tunnelling of the magnetization. The giant spin Hamiltonians derived in our treatment of the dimer clusters contains only (even) powers of the $z$-component of the spin variable. An external magnetic field applied in the $z$ direction can raise and lower the energy levels, causing them to cross at given field strengths. Physical processes that have been omitted from our discussion, such as atomic vibrations, coupling to nuclear spins and transverse magnetic fields, can open up a gap at the level crossings between states with opposite values of the magnetic moment, thereby allowing transitions which represent quantum tunnelling of the magnetization. The Berry curvature will tend to suppress quantum tunnelling by increasing the width of the energy barrier. It is interesting that Berry curvature, which is the curl of the non-adiabatic Mead-Berry potential, and as such is a pure quantum mechanical contribution to the action of the system, suppresses a quantum phenomenon like tunnelling.

The use of the remarkable anisotropy properties of $Co_2$ and $Rh_2$ dimers shown here in information storage memory devices would require positioning them in an environment that preserves, or at least partially preserves, their crucial axial rotational symmetry. We recognize that this is a difficult challenge for nanoscience technologists. However developing nanoscience capabilities which sometimes allow atoms to be manipulated one at the time, for example by STM[23], might make it possible to positioning these dimers on atomic pillars built vertically on a convenient surface. Alternatively one could use the fabrication techniques now employed in molecular-electronic devices, to anchor these dimers inside a nanogap by means of atomically-defined wires. In both cases the axial symmetry could be approximately maintained.

Finally we remark that our predictions for $Co_2$ and $Rh_2$ dimers have implications for the magneto-optical properties of Co and Rh vapours buffered (for example) by inert gas elements. At low temperatures these vapours will act in many ways like a gas phase analogue of ferrofluids. We predict in particular that the Cotton-Mouton effect (the dependence of index of refraction on direction with respect to an external magnetic field) which is proportional[24] to the product of polarizability anisotropy and magnetic susceptibility anisotropy, will be exceptionally strong. The dimer anisotropy energy can be extracted from the temperature dependence of the Cotton-Mouton effect which is sensitive to suppressed magnetic response perpendicular to the molecular axis when the anisotropy is strong.

1. Gatteschi, D., Sessoli, R. & Villain, J. *Molecular Nanomagnets*. (Oxford, New York 2006).

2. Sellmyer, D. & Skomski, R. *Advanced Magnetic Nanostructures*. (Springer, New York 2006)

3. Billas, I. M. L., Châtelain, A., de Heer, W. A. Magnetism from the Atom to the Bulk for Iron, Cobalt and Nickel clusters. *Science* **265**, 1682-1684 (1994).

4. Skomski, R. Nanomagnetics. *J. Phys.: Condens. Matter.* **15**, R841-R896 (2003).

5. Bader, S. D. Colloquium: Opportunities in nanomagnetism. *Rev. Mod. Phys.* **78**, 1 -15 (2006).

6. Gambarella, P. et al. Ferromagnetism in one-dimensional monoatomic metal chains. *Nature* **416**, 301-304 (2002).

7.  Tiago, M. L., Zhou, Y., Alemany, M. M. G., Saad, Y. & Chelikowsky, J. R. Evolution of magnetism in iron from atom to the bulk. *Phys. Rev. Lett*, **97,** 147201:1-4 (2006).

8.  Kresse, G. & Furthmuller, J. Efficiency of ab-initio total energy calculations for metals and semiconductors using a plane-wave basis set. *Comp. Mat. Sci.,* **6**, 15-50 (1996).

9.  Herzberg, G. *Molecular Spectra and Molecular Structure.* (Van Nostrand Reinhold, New York 1950).

10. Jamorski, C., Martinez, A., Castro, M. & Salahub, D. R. Structure of Cobalt clusters up to the tetramer: A density functional study. *Phys. Rev. B*, **55**, 10905-10921 (1997).

11. Castro, M., Jamorski, C. & Salahub, D. R. Structure, bonding, and magnetism of small $Fe_n$, $Co_n$ and $Ni_n$ clusters, n≤5. *Chem. Phys. Lett.* **271**, 133-142 (1997).

12. Gutsev, G.L. & Jr., C. W. B. Chemical bonding, electron affinity, and ionization energies of the homonuclear 3d metal dimers. *J. Phys. Chem. A*, **107**, 4755-4767 (2003).

13. Valiev, M., Bylaska, E. J. & Weare, J. H. Calculations of the electronic structure of the 3d transition metal dimers with projector augmented plane wave method. *J. Chem. Phys.*, **119**, 5955-5964 (2003).

14. Morse, M. D. Clusters of transition-metal atoms. *Chem. Rev.*, **86**, 1049-1109 (1986).

15. Lombardi, J. R. & Davis, B. Periodic properties of force constants of small transition-metal lanthanide clusters. *Chem. Rev.* **102**, 2431-2460 (2002).

16. Cehovin, A., Canali, C. M. & MacDonald, A. H. *Phys. Rev. B*, **66**, 094430-094445 (2002).

17. Canali, C. M., Cehovin, A. & MacDonald, A. H. Chern numbers for spin models of transition metal nanomagnets. *Phys. Rev. Lett.,* **91**, 046805:1-4 (2003).

18. Bohm, A., Mostafadeh, A., Koizumi, H., Niu, Q. & Zwanziger, J. *The Geometric Phase in Quantum Systems.* (Springer Verlag, New York 2003)

19. Auerbach, A. *Interacting electrons and quantum magnetism.* (Springer Verlag, New York 1994).

20. Lazarovits, B., Simon, P., Zarand, G. & Szunyogh, L. Exotic kondo effect from magnetic trimers. *Phys. Rev. Lett.*, 95, 077202:1-4 (2005).

21. Pederson, M. R. & Khanna, S. N. Magnetic anisotropy barrier for spin tunneling in $Mn_{12}O_{12}$ molecules. *Phys. Rev. B*, **60**, 9566-9572 (1999).

22. Garcia-Fernandez, P., Bersuker, I. B. & Boggs, J. E. Lost topological (Berry) phase factor in electronic structure calculations. Example: the ozone molecule. *Phys. Rev. Lett.,* **96**, 163005:1-4 (2006).

23. C.F. Hirjibehedin, C.P. Lutz and A.J. Heinrich, "Spin coupling in engineered atomic structures", Science, **312,** 1021-1024 (2006).

24. Rizzo,C., Rizzo, A. & Bishop, D. M.  The Cotton-Mouton effect in gases: experiment and theory. *Int. Rev. in Phys. Chem.*,  **16,** 81-111 (1997).

| System | GS | J |
|--------|-----|---|
| $Co_2$ | $\Delta(S=2)_g$ | 4 |
| $Rh_2$ | $\Delta(S=2)_g$ | 4 |
| $Ni_2$ | $\Sigma(S=1)_g$ | 1 |
| $Pd_2$ | $\Sigma(S=1)_g$ | 1 |
| $Fe_2$ | $\Delta(S=3)_g$ | 3 |
| $Ru_2$ | $\Sigma(S=2)_g$ | 2 |

Table 1: **Classification of the electronic ground state for 3d and 4d transition metal dimers.** In the second column, the molecular GS in the absence of spin-orbit coupling, characterized by total $L_z$, total spin $S$ and parity ($g$ or $u$). In the third column, the Chern numbers $J$ with spin-orbit interaction are listed. The Chern number gives the dimension of the effective giant-spin Hilbert space. Note that when the HOMO is non-degenerate, $J=S$.

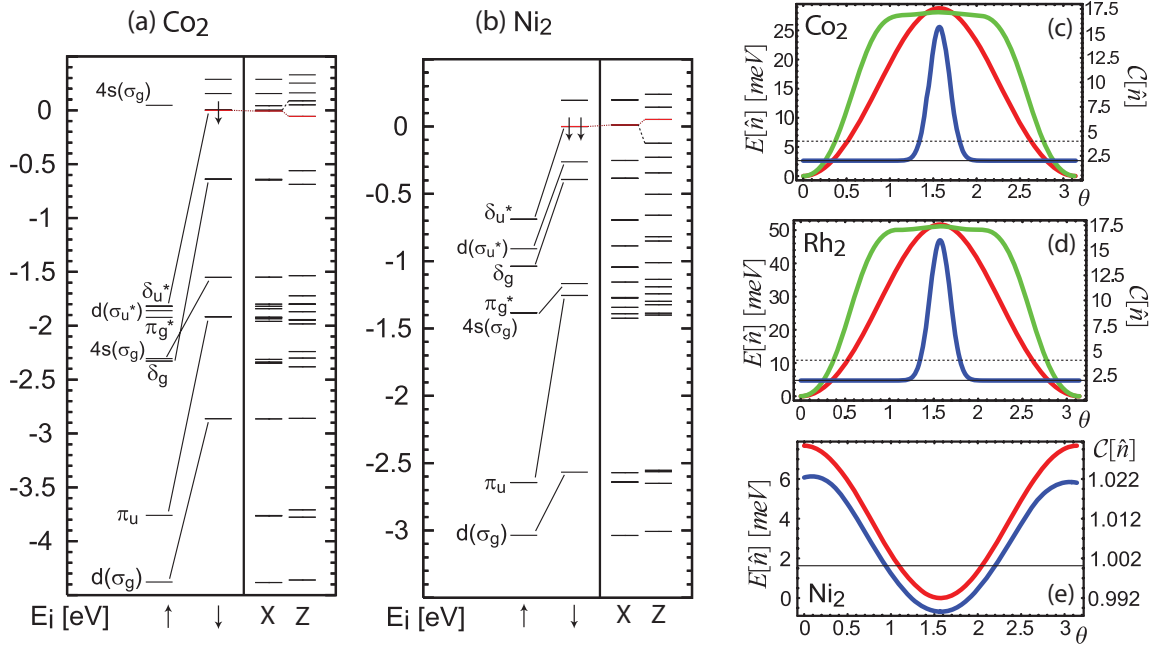

Figure 1: Electronic molecular orbital energies for (a) Co$_2$ and (b) Ni$_2$. The left part of each panel represents the spectrum in the absence of spin-orbit interaction, where each level is either minority (↓) or majority (↑) spin. The energy of the HOMO (in red) is taken as a reference and set equal to zero. The right part of the panel represents the spectrum when spin-orbit coupling is present, calculated for two directions of the magnetic moment: $\hat{z}$ is along the dimer axis and $\hat{x}$ is along a direction in plane perpendicular to the dimer axis. The three graphs on the right show the magnetic anisotropy $E[\hat{n}]$ (red) and Berry's curvature $C[\hat{n}]$ (blue) for (c) Co$_2$, (d) Rh$_2$ and (e) Ni$_2$ as a function of the angle $\theta$ between the magnetic moment direction $\hat{n}$ and the dimer axis. Both quantities are symmetric around the dimer axis. The green line represents the magnetic anisotropy energy corrected by the Berry curvature, as obtained from the effective giant-spin Hamiltonian $\mathcal{H}[J, \hat{n}]$. In Ni$_2$ $C[\hat{n}]$ is a smooth function of $\hat{n}$ because of the relatively large energy gap between the HOMO and the LUMO and the effective giant-spin Hamiltonian is indistinguishable from $E[\hat{n}]$.

Acknowledgements: We would like to thank David Bishop, Walt de Heer and John Keto for helpful discussions. This work was supported in part by the Welch

Foundation, the National Science Foundation under grant DMR-0606489. the Faculty of Natural Sciences at Kalmar University, the Swedish Research Council under Grant No: 621-2004-4439, and by the Office of Naval Research.

Correspondence: Correspondence and requests for materials should be addressed to T. O. Strandberg (olof.strandberg@hik.se)